\documentclass[12pt,a4paper]{article}
\usepackage[ansinew]{inputenc}
\usepackage{amsthm}
\usepackage{amssymb,bm}
\usepackage{amsmath}
\usepackage[english]{babel}
\usepackage{array}
\usepackage{amsfonts}
\usepackage{latexsym}
\usepackage{longtable}
\numberwithin{figure}{section}
\numberwithin{table}{section}

\usepackage[pdftex]{color,graphicx}
\usepackage[usenames,dvipsnames,table]{xcolor}
\usepackage{float} 
\usepackage{subfigure} 
\usepackage{wrapfig} 
\usepackage[rflt]{floatflt} 

\date{}

\numberwithin{equation}{section}
\numberwithin{figure}{section}

\def \II {\hbox{\boldmath$I$}}

\def \VV {\hbox{\boldmath$V$}}
\def \XX {\hbox{\boldmath$X$}}

\def \ZZ {\hbox{\boldmath$Z$}}

\def \ee {\hbox{\boldmath$e$}}

\def \uu {\hbox{\boldmath$u$}}

\def \xx {\hbox{\boldmath$x$}}
\def \yy {\hbox{\boldmath$y$}}

\def \bbeta {\hbox{\boldmath$\beta$}}

\title{\textbf{Prevalence of international migration: an alternative for small area estimation}}
\author{Jairo F\'uquene$^{a}$, Cesar Cristancho$^{b}$, Mariana Ospina$^{b}$, Domingo Morales$^{c}$. \vspace{0.5 cm} \\
{\small  $^{a}$ Technical Consultant for National Administrative Department of Statistics, Colombia.}\\
{\small  $^{b}$ Population Projections Division. National Administrative Department of Statistics, Colombia.}\\
{\small  $^{c}$ Operations Research Center, University Miguel Hern\'{a}ndez de Elche, Spain.}}

\begin{document}

\maketitle

\date

\noindent\textbf{Abstract}
\medskip

\noindent
This paper introduces an alternative procedure for estimating the prevalence of international migration at the municipal level in Colombia. The new methodology uses the empirical best linear unbiased predictor based on a Fay-Herriot model with target and auxiliary variables available from census studies and from the Demographic and Health Survey. The proposed alternative produces prevalence estimates which are consistent with sample sizes and demographic dynamics in Colombia. Additionally, the estimated coefficients of variation are lower than 20\% for municipalities and large demographically-relevant capital cities and therefore estimates may be considered as reliable.
\vspace*{0.1 cm}

\section{Introduction}
According with data recently released by Migración-Colombia (2015) the migratory flow of Colombians abroad in the period 2007–2015 displayed an upward trend. The registered data reflect the upward trends in migration prevalence and the variability over the years. For instance, in 2012 Colombia showed the greatest increase in migration abroad
(approximately 25\%) with respect to 2011. Migration is a key element for growth (and downturn) in populations. Edmonston and Michalowski (2004) showed that changes in the demographic characteristics of origin and destination areas are affected by migration flows and hence the importance of having estimations that give an account of the characteristics for this phenomenon.

In the case of Colombia, there is a marked contrast between the growing need for disaggregated data at a local level for planning purposes and the lack of adequate sources, tools and research with the level of detail required. Despite the fact that internal migration has a greater impact than international migration in terms of population growth and demographic change in the majority of municipalities, there is a growing need to research the effects of international migration as it has a significant impact on the demographic dynamics of some of the areas of interest. See Smith et al. (2013) for a practical guide on state and local population projections.

An important source of information to measure migration prevalence (proportion of individuals in a population which displays the migratory event in a given moment or period of time) abroad at the municipal level is the population census carried out in 2005 by the National Administrative Department of Statistics (DANE, by its Spanish acronym). Another recent information source is the Demographic and Health Survey of 2015 (DHS2015), in which it is possible to measure the number of households that have experienced migration abroad at a departmental level. See Profamilia (2015) for a description of DHS2015.

It is possible to make estimations of international migration prevalences for 2015 at the municipal level by borrowing strength of prevalences obtained from the 2005 census, which can display considerable bias if they are taken as linear trends over the years. For this reason, it is essential for the country to be able to measure migration prevalence at the municipal level with greater precision. A methodological proposal to estimate migration prevalences at the municipal level would make it possible to have estimations for intercensal periods and, additionally, would help to update estimations on the level and structure of population dynamics in Colombia. Likewise, a migration study at the municipal level would help to illustrate the health, education and employment requirements of the new population.

The goal of this paper is to explore for the first time (for the best of our knowledge), the use of small area estimation (SAE) theory to estimate the prevalence of migration abroad in intercensal periods at the municipal level. Using the Colombian demographic context as a reference, the paper also establishes whether the prevalence estimates based on the proposed methodology are reliable and in keeping with population dynamics. Therefore, this work proposes an alternative methodological framework to estimate the prevalence of migration abroad at the municipal level in Colombia.

Statistical models have helped researchers to estimate migration flows between regions or countries. For example, Willekens (2008) applied unit-level models to calculate the probability that individuals in the population will migrate. At the aggregated level, Plane (1982) and Raymer (2007) proposed gravity, entropy maximization and log-linear models to estimate aggregate migration flows between countries and Raymer and Rogers (2007) incorporated spatial dependence and socio-economic similarities between origin and destination countries. However, these approaches do not consider the problem of using survey data to estimate migration in regions where the amount available information is small.

The sampling and inference techniques for finite populations are used to calculate direct estimators of international migration prevalences and their variances. These direct estimates are entered into a Fay-Herriot (FH) model that enables the calculation of empirical best linear unbiased predictors (EBLUP) of international migration prevalences ate the municipal level.

The Fay-Herriot model is a linear mixed model proposed by Fay and Herriot (1979) to estimate the average per capita income in small areas in the United States. Esteban et al. (2012) and Morales et al. (2015), Benavent and Morales (2016) and Burgard et al. (2019) extended this methodology to the estimation of small area poverty proportions under temporal, bivariate and measurement error Fay-Herriot models respectively, but they did not give applications to demography or to population dynamics. Rao and Molina (2015) gave a complete introduction to the theory and practice of small area estimation.

The rest of this paper is organized into three sections. Section \ref{ssec:dos} introduces the Fay-Herriot model and presents the estimators of migration prevalences. This section also provides an adequate modeling of the estimated design-based variances of the direct estimators. Section 3 applies the methodology of Section 2 to estimate the prevalence of migration abroad in municipalities of Colombia. Additionally, Section \ref{ssec:tres} carries out a model diagnosis along with a comparison with traditional estimators. Finally, Section \ref{ssec:cuatro} presents some conclusions related to the obtained results.
\section{Migration prevalence estimation under the Fay-Herriot model}\label{ssec:dos}
Consider a finite population of size $N$, where the units are households. For each household $j$, the variable of interest is $y_{j} =1$ if at least a person in the household $j$ currently lives abroad, and it is $y_{j} =0$ if no one in the household $j$ currently lives abroad. Assume that the population is divided into small areas labelled by $d = 1,\ldots,D$. In our case, they are the municipalities of Colombia. Let $s$ be a sample extracted at random from the population according to a given sample design. The population and the sample are denoted by $U=\cup_{d=1}^{D}U_{d}$ and  $s=\cup_{d=1}^{D}s_{d}$, respectively. Let $N_d$ and $n_d$, be the sizes of $U_d$ and $s_d$, $d=1,\ldots,D$.
The objective is to estimate the prevalence of migration abroad per municipality, $\overline{Y}_{d}=N_d^{-1}\sum_{j \in U_{d}}y_{j}$, with its respective coefficient of variation. The direct estimator of $\overline{Y}_{d}$, proposed by Hájek (1971), and the corresponding estimated coefficient of variation  are
\begin{equation}\label{eq:directo}
\hat{\bar{Y}}_{d}^{dir}=\frac{1}{\hat{N}_{d}}\sum_{j \in s_{d}} {w}_{j} y_{j},\quad
\hat{\text{cv}}^{dir}\big(\hat{\bar{Y}}_{d}^{dir}\big)=
\frac{1}{\hat{\bar{Y}}_{d}^{dir}}
\sqrt{\dfrac{1}{\hat{N}_{d}^{2}}\sum_{j \in s_{d}}{w}_{j}({w}_{j}-1)
\big(y_{j}-\hat{\bar{Y}}_{d}^{dir}\big)^2},
\end{equation}
where $\hat{N}_{d}=\sum_{j \in s_{d}} {w}_{j}$ is the direct estimator of the population size of municipality $d$ and ${w}_{j}$ is the sample weight or expansion factor of household $j$. The sample weights of the DHS2015 survey are obtained from a complex sampling design in four stages: (1) municipalities, (2) urban blocks and rural areas, (3) segments and, (4) people. The quartiles of the sample sizes of the $D=282$ municipalities represented in the DHS2015 survey are $n^{(0)}=7$, $n^{(1)}=47$, $n^{(2)}=89$, $n^{(3)}=170$, $n^{(4)}=3069$. Therefore, estimating municipality indicators is a SAE problem.

The direct estimators are approximately unbiased for $\overline{Y}_{d}$, $d=1,\ldots,D$, but they are not usually precise for estimating small areas parameters. Because of the small sample sizes, they lead to considerably high coefficients of variation ($>20\%$).
See S\"arndal et al. (1992) for more details about inference in finite population and properties of direct estimators.

Small area estimators might be based on unit-level or area-level models.
Under the area-level approach, a model is fitted to the aggregated data $(\hat{\bar{Y}}_{d}^{dir},\xx_{d})$, $d=1,\ldots,D$, where $\xx_{d}$ is a row vector that contains the values of $p$ auxiliary variables. The use of area-level SAE procedures has the disadvantage of using auxiliary variables aggregated at the municipality level, with the possible loss of information with respect to the data aggregated at a lower level (for example, blocks, segments or people). However, this disadvantage is offset by using all the auxiliary variables available at the municipal level.
This fact also prevents the restriction of estimators based on unit-level models that must have the same auxiliary variables in the survey file and in the external administrative records.
\subsection{The Fay-Herriot model}
Fay and Herriot (1979) introduced an area-level linear mixed model for small area estimation.
The Fay-Herriot model for estimating the prevalence of international migration can be expressed in the matrix form
\begin{equation}\label{eq:FH}
\yy= \XX\bbeta + \ZZ\uu + \ee,
\end{equation}
where $\yy=(\hat{\bar{Y}}_{1}^{dir},\ldots,\hat{\bar{Y}}_{D}^{dir})^\prime$ is the vector that contains the direct prevalence estimates, $\XX=\text{col}_{1\leq d \leq D}(\xx_{d})$ is the $D\times p$ matrix containing the values of $p$ auxiliary variables, $\bbeta=(\beta_{1},\ldots,\beta_{D})^\prime$ is the vector of regression parameters, $\ZZ=\II_{D}$ and $\II_D$ is the $D\times D$ identity matrix.
Concerning the random effects and errors, the model assumes that $\uu=(u_{1},\ldots,u_{D})^\prime\sim N(\boldsymbol{0},\sigma^{2}_{u}\II_{D})$ and
$\ee=(e_{1},\ldots,e_{D})^\prime\sim N(\boldsymbol{0},\VV_e)$ are independent, where $\VV_e=\text{diag}_{1\leq d\leq D}(\sigma_d^2)$ and the variances $\sigma_d^2$ are assumed to be known.
In practice, the error variances $\sigma_d^2$ can be substituted by estimates $\hat{\sigma}_d^2$ calculated from unit-level data.
In any case, the Fay-Herriot treats these estimates as known constants.

There are several method for estimating the parameters of model (\ref{eq:FH}), like maximum likelihood (ML), residual maximum likelihood (REML) or method of moments.
This paper applies the REML method to calculate an estimator $\hat\sigma_u^2$ of the random effect variance.
The REML estimator of $\sigma_u^2$ has lower bias that its ML counterpart.
Further the REML method avoids potential negative variance estimates such as in the case of the method of moments.
The REML estimator of the vector of regression parameters and the EBLUP of the random effects are
\begin{equation}\label{eq:MLE}
\hat{\bbeta}=(\XX^\prime\hat{\VV}^{-1}\XX)^{-1}\XX^\prime\hat{\VV}^{-1}\yy,\quad
\hat{u}_{d}=\dfrac{\hat{\sigma}^{2}_{u}}{\hat{\sigma}^{2}_{u}+\hat{\sigma}^{2}_{d}}(\hat{\bar{Y}}_{d}^{dir}-\xx_{d}\hat{\bbeta}),\quad d=1,\ldots,D.
\end{equation}
where $\hat{\VV}=\text{diag}(\hat{\sigma}^{2}_{u}+\hat{\sigma}^{2}_{d},\ldots,\hat{\sigma}^{2}_{u}+\hat{\sigma}^{2}_{D})$.
Under the Fay-Herriot model (\ref{eq:FH}), the EBLUP of $\bar{Y}_d$ is
\begin{equation}\label{eq:EBLUP}
\hat{\bar{Y}}_{d}^{FH}=\frac{\hat{\sigma}^{2}_{u}}{\hat{\sigma}^{2}_{u}+\hat{\sigma}^{2}_{d}}\,\hat{\bar{Y}}_{d}^{dir}
+\frac{\hat{\sigma}^{2}_{d}}{\hat{\sigma}^{2}_{u}+\hat{\sigma}^{2}_{d}}\,
\xx_{d}\hat{\bbeta},\quad d=1,\ldots,D.
\end{equation}
The EBLUP (\ref{eq:EBLUP}) is a weighted average of the estimate produced by the direct estimator and the values predicted by model (\ref{eq:FH}). If the sample size per municipality $n_{d}$ is large, it is understood that $\hat{\sigma}^{2}_{d}\approx 0$ and therefore the EBLUP and the direct estimator are approximately equal, i.e., $\hat{\bar{Y}}_{d}^{FH}\approx
\hat{\bar{Y}}_{d}^{dir}$. Conversely, if in the calculation of the direct estimator, there is a small sample size per municipality then $\hat{\sigma}^{2}_{d}>\hat{\sigma}^{2}_{u}$ and therefore the EBLUP is approximately equal to the values predicted in the Fay-Herriot model, i.e., $\hat{\bar{Y}}_{d}^{FH}\approx
\xx_{d}\hat{\bbeta}$.

For calculating the predictor (\ref{eq:EBLUP}) it is necessary to estimate the $\sigma^{2}_{d}$  variances. Using unit-level data and aplying the formula (\ref{eq:directo}), it is possible to obtain direct estimates, $\hat{\text{var}}(\hat{\bar{Y}}_{d}^{dir})$, of the variance of the direct estimator $\hat{\bar{Y}}_{d}^{dir}$. However, these estimates are not precise since the sample sizes of the municipalities are generally very small. An alternative estimation procedure uses auxiliary data and applies a log-linear model for the variance estimates of direct estimators. This is the method of the Generalized Variance Function (GVF), which is described, for example, in Schall (1991). The GVF method fits the model
\begin{equation}\label{eq:loglineal}
\log(\hat{\text{var}}(\hat{\bar{Y}}_{d}^{dir}))=\xx^{v}_{d}\bbeta^{v} + \varepsilon^{v}_{d},\quad d=1,\ldots,D,
\end{equation}
where $\varepsilon_d^{dir}\sim N(0,\sigma_\varepsilon^2)$, $d=1,\ldots,D$, are independent. In order to estimate $\sigma^{2}_{d}$, the predicted values are obtained from model (\ref{eq:loglineal}) using the following formula
\begin{equation}\label{eq:predsigma}
\hat{\sigma}^{2}_{d}=\exp\{\hat{\text{var}}(\hat{\bar{Y}}_{d}^{dir})/2\}
\exp\{\xx^{v}_{d}\hat{\bbeta}^{v}\},\quad d=1,\ldots,D,
\end{equation}
where $\exp\{\hat{\text{var}}(\hat{\bar{Y}}_{d}^{dir})\}$
is the bias-correction term in log-normal regression. This paper selects the auxiliary variables contained in the vector $\xx^{v}_{d}$ from the direct estimates of migration prevalences, $\hat{\bar{Y}}_{d}^{dir}$, the sample sizes $n_{d}$ and the interaction between them. For improving the predictive capacity of the GVF model (\ref{eq:loglineal}), the included covariates can be raised to an exponent resolving potential problems of heteroscedasticity. In practice, when the correction term is not used, the GVF method tends to underestimate the variances.

To estimate the coefficient of variation of the EBLUP $\hat{\bar{Y}}_{d}^{FH}$, we approximate its the mean squared error (MSE).
Prasad and Rao (1990) gave the MSE estimator
\begin{equation}\label{eq:MSE}
\text{MSE}(\hat{\bar{Y}}_{d}^{FH})=
g_{1}(\hat{\sigma}^{2}_{u})+
g_{2}(\hat{\sigma}^{2}_{u})+
2g_{3}(\hat{\sigma}^{2}_{u}),
\end{equation}
where
\begin{align*}
g_{1}(\hat{\sigma}^{2}_{u})&=\dfrac{\hat{\sigma}^{2}_{u}\hat{\sigma}^{2}_{d}}{
\hat{\sigma}^{2}_{u}+\hat{\sigma}^{2}_{d}}, &
g_{2}(\hat{\sigma}^{2}_{u})&=\dfrac{\hat{\sigma}^{4}_{d}}{
(\hat{\sigma}^{2}_{u}+\hat{\sigma}^{2}_{d})^{2}}(\XX^\prime\hat{\VV}^{-1}\XX)^{-1}\xx_{d}^\prime, &
g_{3}(\hat{\sigma}^{2}_{u})&=\dfrac{\hat{\sigma}^{4}_{d}}{
(\hat{\sigma}^{2}_{u}+\hat{\sigma}^{2}_{d})^{3}}\text{avar}(\hat{\sigma}^{2}_{u}),
\end{align*}
with $g_{3}(\hat{\sigma}^{2}_{u})$ given by Datta and Lahiri (2000) and where the asymptotic variance is
\begin{equation}
\text{avar}(\hat{\sigma}^{2}_{u})=2\bigg(\sum_{d=1}^{D}
1/(\hat{\sigma}^{2}_{u}+\hat{\sigma}^{2}_{d})^{2}
\bigg)^{-1}.
\end{equation}
\section{Migration prevalence estimation in Colombia}\label{ssec:tres}
This section fits a Fay-Herriot model to DHS2015 data and calculates the EBLUP of the prevalence of migration abroad at the municipal level. Section \ref{subssec:model} applies the GVF method to estimate the variances of the direct estimators of migration prevalences. The final variance estimates are the predicted values of a selected log-linear regression model. Section \ref{subssec:diagnosis} presents the comparisons between the EBLUPs and the direct estimators of migration prevalences and the corresponding diagnosis of the fitted Fay-Herriot model.
\subsection{Modeling the variance of the direct estimator}\label{subssec:model}

As mentioned in Section \ref{ssec:dos}, in order to improve the application of the Fay-Herriot model a GVF estimation of the variance of the direct estimator is carried out. Since, in some municipalities the direct estimate and its variance equals zero a term $\delta$ is added to the variance of the direct estimator in model (\ref{eq:loglineal}) to prevent numerical errors and/or heteroscedasticity in the residuals. However, due to the term $\delta$, the correction $\exp\{\hat{\text{var}}(\hat{\bar{Y}}_{d}^{dir})\}$ is not applied in (\ref{eq:predsigma}). In practice, there are different possibilities for the selection of the value $\delta$. In this work, we take $\delta$ as the average of the direct estimates of the design-based variances, which is $\delta=0.0056$.

For the GVF model (\ref{eq:loglineal}), the selected explicative variables and parameters are
\begin{equation}\label{eq:loglineal2}
\xx_d^{v}=\big(1,\hat{\bar{Y}}_{d}^{dir},
\big(\hat{\bar{Y}}_{d}^{dir}\big)^{1/2},n_{d}, n_{d}^{1/2},\big(\hat{\bar{Y}}_{d}^{dir}n_{d}\big)^{1/2}\big),\quad
\hat{\bbeta}^{v}=\big(\beta_{1}^{v}, \beta_{2}^{v}, \beta_{3}^{v}, \beta_{4}^{v},\beta_{5}^{v},\beta_{6}^{v}\big)^\prime.
\end{equation}
The selection of the variables of the log-linear model is made through the Akaike Information Criterium (AIC) in which the final model includes all the variables proposed in (\ref{eq:loglineal2}).
Table \ref{tab:residuales} presents the estimates of the regression parameters of the selected GVF model.
They are all significant with $p$-values lower than 0.001.

\begin{table}[ht]
\centering
\renewcommand{\arraystretch}{1.0}
\begin{tabular}{rrrrr}
  \hline
& estimate	& standard error	& $t$-value	& $p$-value \\
  \hline
$\beta_{1}^{v}$ & -7.3518 & 0.0344 & -213.94 & $<0.001$ \\
$\beta_{2}^{v}$ & 6.3722 & 0.9335 & 6.83 & $<0.001$ \\
$\beta_{3}^{v}$ & 5.2049 & 0.2898 & 17.96 & $<0.001$ \\
$\beta_{4}^{v}$ & 0.0014 & 0.0001 & 14.47 & $<0.001$ \\
$\beta_{5}^{v}$ & -0.0325 & 0.0045 & -7.17 & $<0.001$ \\
$\beta_{6}^{v}$ & -0.3017 & 0.0196 & -15.40 & $<0.001$ \\
   \hline
\end{tabular}

\vspace*{-1mm}
\caption{Estimated parameters of the GVF model.}
\label{tab:residuales}
\end{table}
The signs of the regression parameters basically say that the variance of the direct estimator of the migration prevalence increases with the estimated prevalence and decreases with the sample size.
It is interesting to observe that $\beta_{6}^{v}<0$, so that the direct estimator and the sample sizes interact negatively.

\begin{figure}[h]
\begin{center}
\includegraphics[scale=0.58]{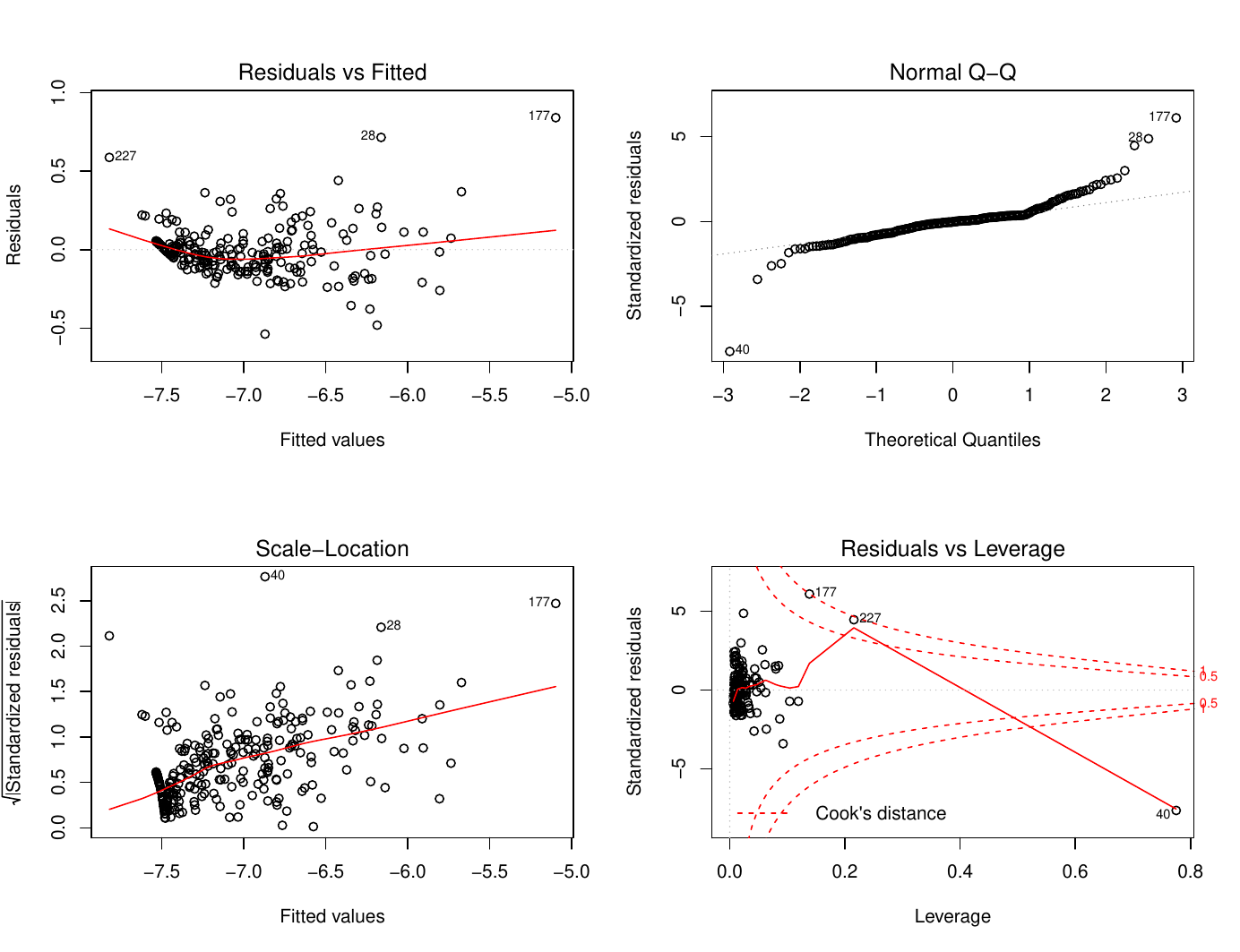}
\end{center}

\vspace*{-6mm}
\caption{Residuals of the selected GVF model.}
\label{fig:residuales2}
\end{figure}

Figure \ref{fig:residuales2} plots the residuals of model (\ref{eq:loglineal}) with the explicative variables and parameters (\ref{eq:loglineal2}). The residuals do not present problems of heteroscedasticity or normality and no significant outliers are observed.

\begin{figure}[ht!]
\begin{center}
\includegraphics[width=0.42\textwidth]{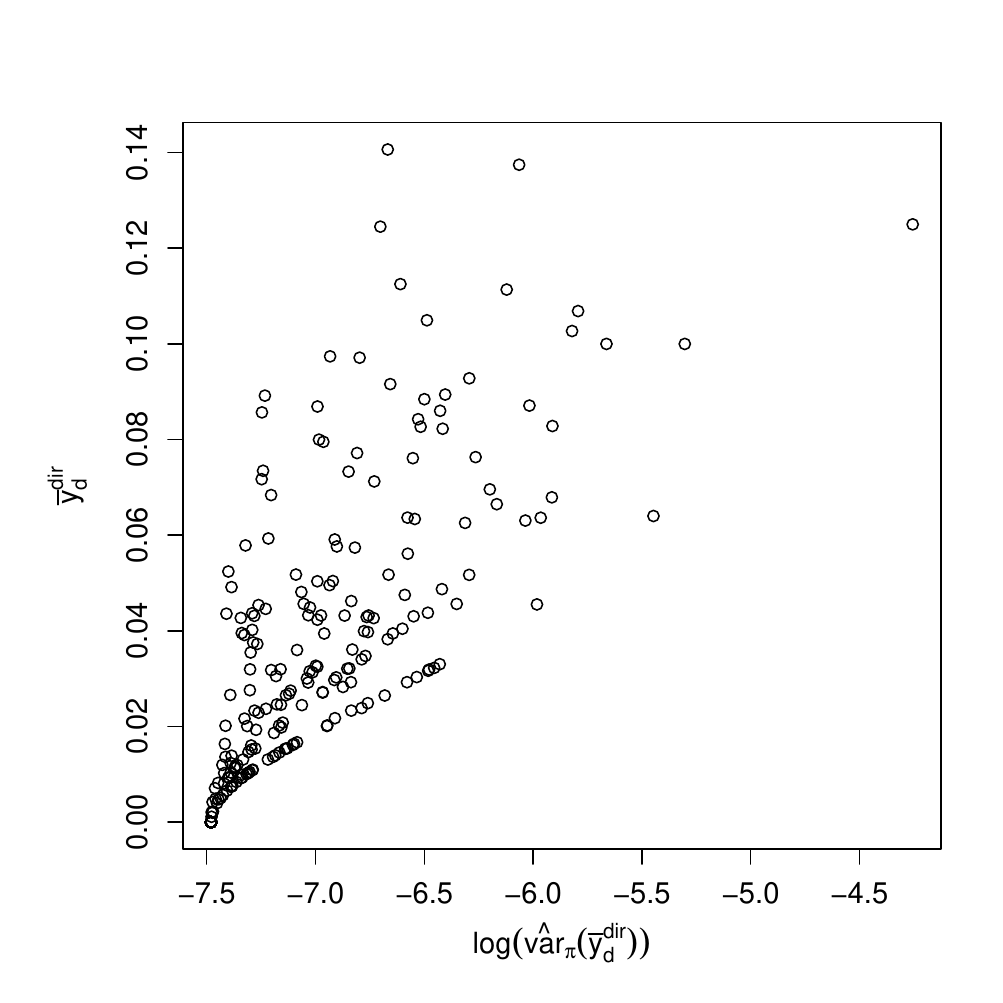}\quad
\includegraphics[width=0.42\textwidth]{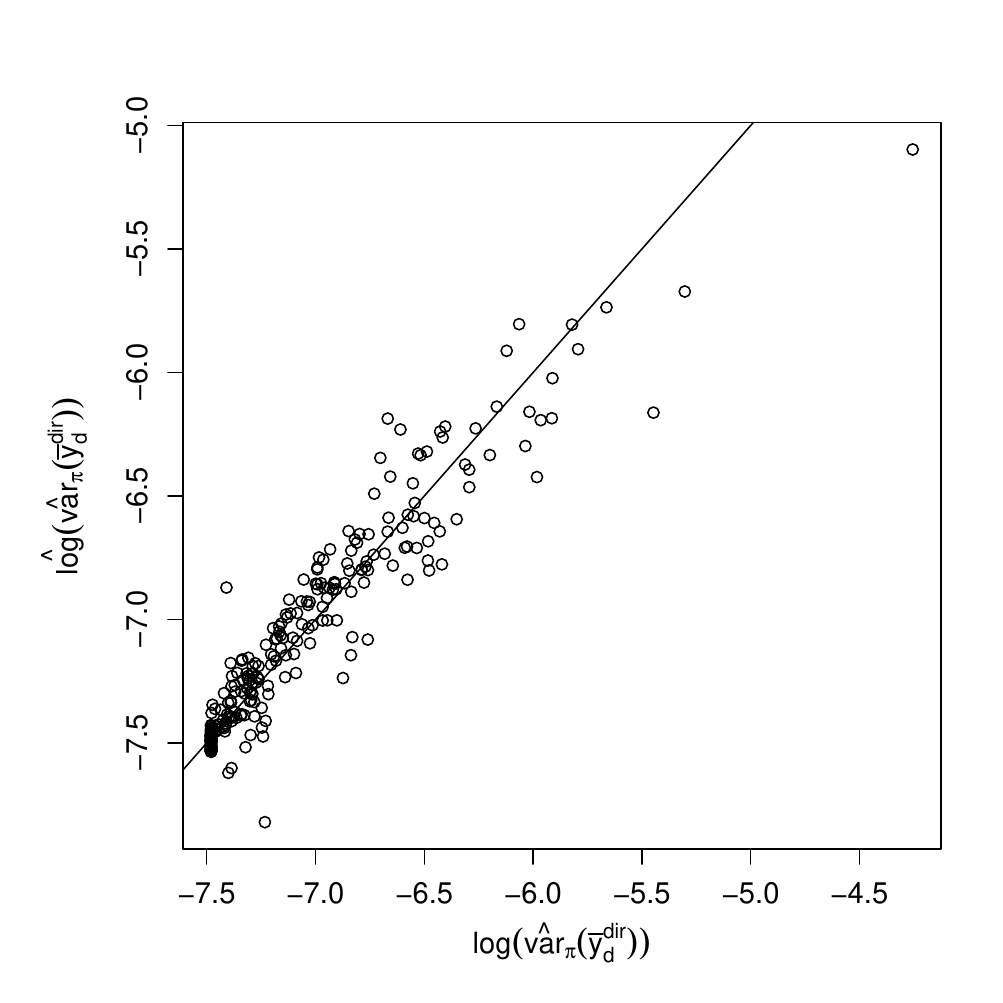}
\end{center}

\vspace*{-6mm}
\caption{Dispersion graphs for observed values under the fitted GVF model.}
\label{fig:all1}
\end{figure}

Figure \ref{fig:all1} (left) plots a dispersion graph between the direct estimates and the logarithms of their estimated variances.
Figure \ref{fig:all1} (right) plots a dispersion graph between the predicted logarithms of the estimated variances and the logarithms of the estimated variances; this is to say, between the predicted and observed values of the GVF model.
In both cases, we observe a positive linear relationship that validates the selected GVF model.

\begin{figure}[ht]
\begin{center}
\includegraphics[width=0.45\textwidth]{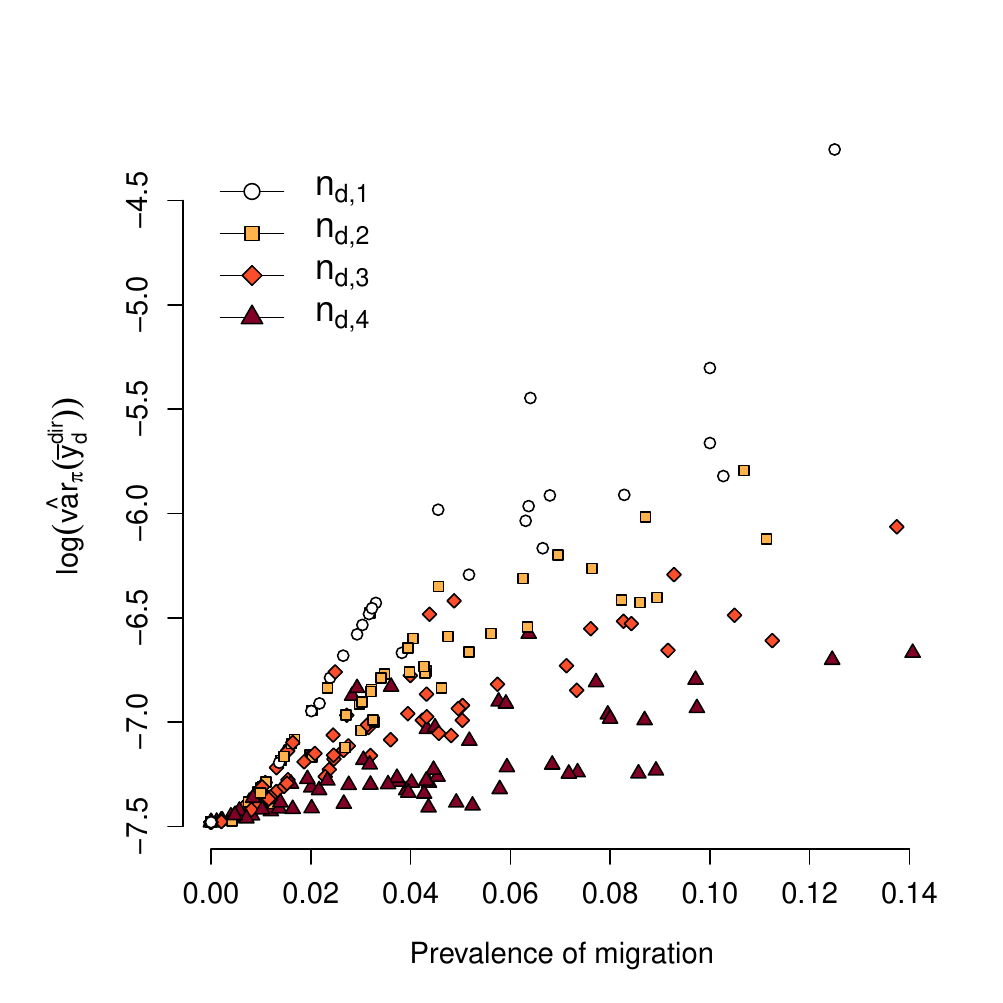}\quad
\includegraphics[width=0.45\textwidth]{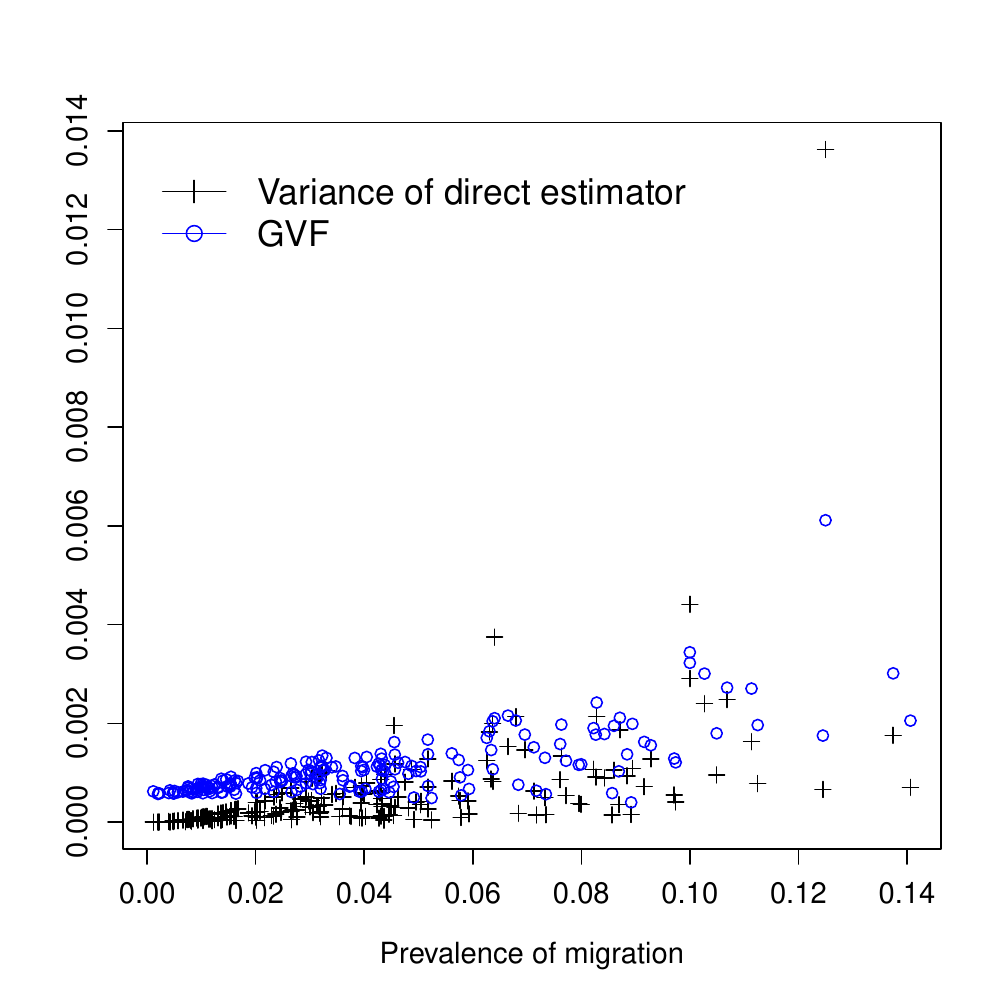}
\end{center}

\vspace*{-6mm}
\caption{Dispersion graphs for prevalence of migration under the fitted GVF model.}
\label{fig:all2}
\end{figure}

To construct Figure \ref{fig:all2} (left), we partition the set of sample sizes into four subsets using the quartiles as cut points. From lowest to highest, sets are denoted by $n_{d,1},...,n_{d,4}$. Figure \ref{fig:all2} (left) illustrates how the log-variance of the direct estimator decreases when the sample size increases, which is natural from a practical point of view. In Figure \ref{fig:all2} (right), it is possible to observe how the variance of the direct estimator and its prediction increase with similar behavior for different values in the estimation of migration prevalence.

Because of the results of the diagnosis analysis, we take the predicted GVF variances as true error variances for the Fay-Herrot model on the direct estimates of international prevalences.
In this case there is no need of applying a nonparametric variance estimation procedure, like the one suggested by Gonz\'alez-Manteiga at al. (2010).

\subsection{EBLUPs of international migration prevalences}\label{subssec:diagnosis}
This section fits a Fay-Herriot model to the estimates of the prevalence of international migration in municipalities of Colombia, performs a diagnosis analysis and presents the results of the employed methodology.
For the Fay-Herriot model introduced in (\ref{eq:FH}), the use of covariates that provide the greatest possible degree of information at the municipal level is highly important. One of the objectives of this work is to estimate the migration prevalence in intercensal periods. For this reason, we use data from the census carried out right before the survey DHS2015 for introducing the auxiliary variables of the selected Fay-Herriot model.
The first covariate, $x_1$, is the small area average of the unit-level variable
$$
x_{1j} =
  \begin{cases}
    1 & \text{if any person in the household $j$ lived abroad during 2005,}  \\
    0 & \text{if no one in the household $j$ lived abroad during 2005,}
  \end{cases}
$$
for $j=1,\ldots,N_{d}$. This variable is taken from the 2005 census.
The second covariable, $x_{2}$, is the multidimensional poverty index described by Anand and Sen (1997) and expressed as a percentage. This variable is calculated using the methodology provided by the Oxford Poverty and Human Development Initiative (OPHI) and is constructed with variables obtained from the 2005 Census.

The Fay-Herriot model that includes only the variable $x_{2}$ has AIC=-1252.266 and the model with the variables $x_{1}$ and $x_{2}$ has AIC=-1311.098. The second model has lower AIC, a better interpretation of the results and an acceptable analysis of residuals. For these reasons, the second model is selected.
Table \ref{tab:residuales2} presents the estimated parameters of the selected model.

\begin{table}[ht]
\renewcommand{\arraystretch}{1.0}
\begin{center}
\begin{tabular}{rrrrr}
  \hline
& estimate	& standard error	& $t$-value	& $p$-value \\
  \hline
$\beta_{0}$ & 0.02 & 0.01 & 3.10 & $<0.001$ \\
$\beta_{1}$ & 1.63 & 0.21 & 7.80 & $<0.001$ \\
$\beta_{2}$ & -0.02 & 0.01 & -1.81 & $<0.1$ \\
   \hline
\end{tabular}

\vspace*{-1mm}
\caption{Estimated parameters of the selected Fay-Herriot model.}
\label{tab:residuales2}
\end{center}
\end{table}

Table \ref{tab:residuales2} shows how the parameters are significant with $p$-values lower than 0.1 and they indicate that as the prevalence of international migration increases in 2005, the prevalence of migration increases in 2015 and as the poverty index decreases the experience of migration increases.

\begin{figure}[ht]
\begin{center}
\includegraphics[width=0.42\textwidth]{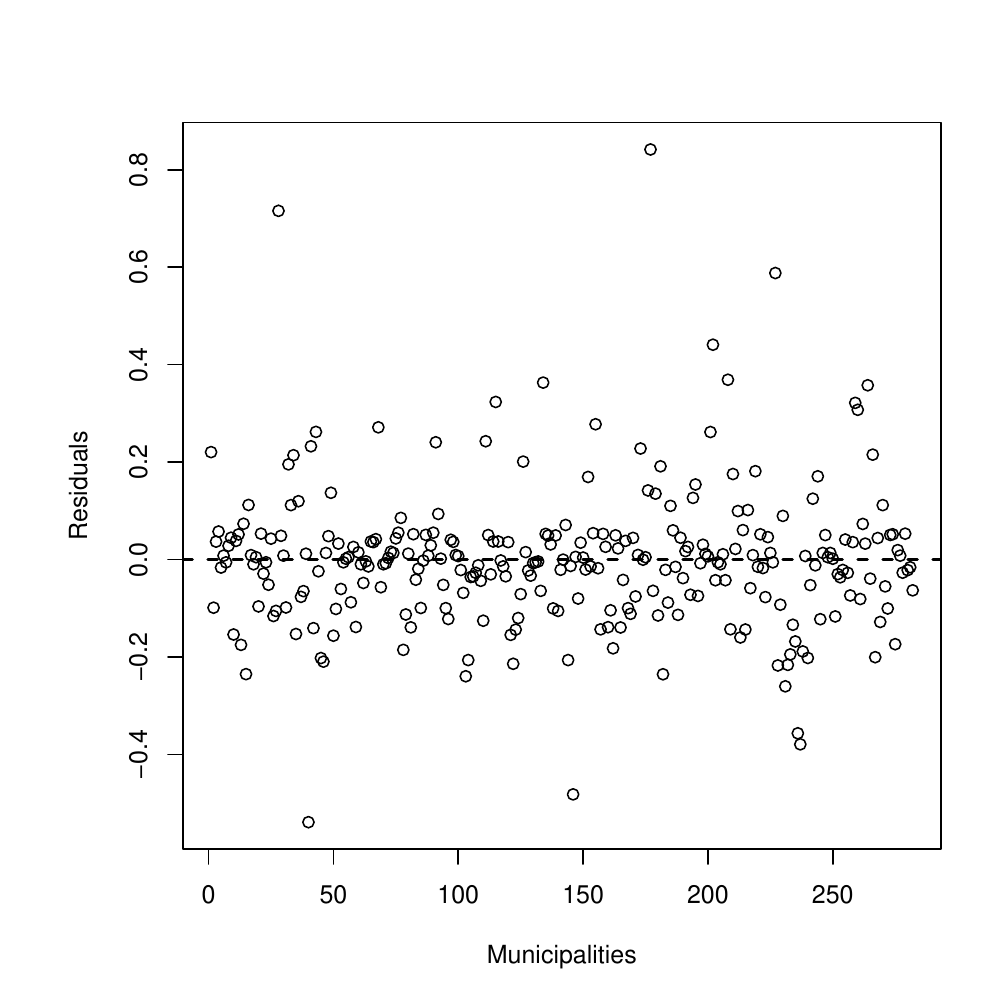}
\includegraphics[width=0.42\textwidth]{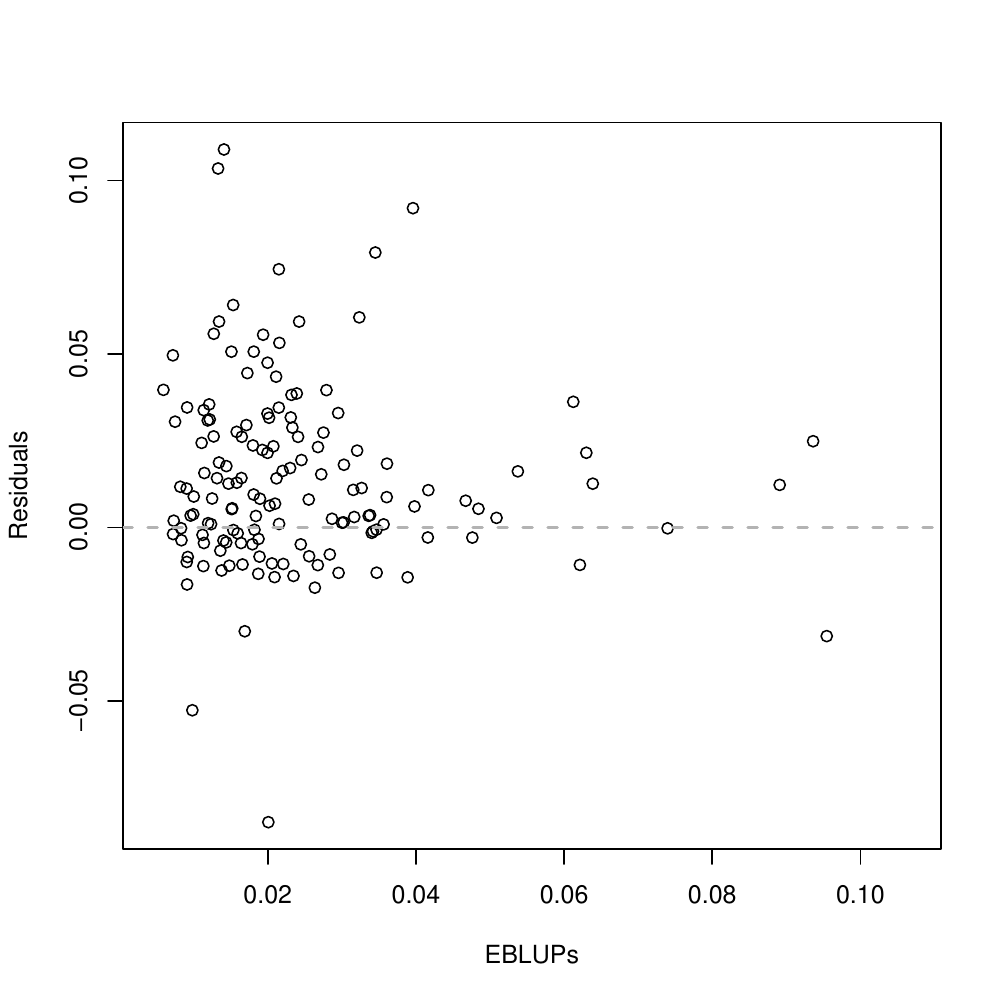}
\end{center}

\vspace*{-6mm}
\caption{Residuals of the selected Fay-Herriot model.}
\label{fig:resiFH}
\end{figure}

The residuals of the Fay-Herriot model are computed as the different between the estimators, i.e.
$\hat{e}_d=\hat{\bar{Y}}_d^{\text{dir}}-\hat{\bar{Y}}_d^{\text{FH}}$, $d=1,\ldots,D$.
Figure \ref{fig:resiFH} (left) shows how the residuals of the Fay-Herriot model are random without any evident correlation between municipalities. Figure \ref{fig:resiFH} (right) illustrates how the randomness of residuals is maintained through the EBLUPs of the prevalences.

\begin{figure}[ht!]
\begin{center}
\vspace*{-6mm}

\includegraphics[scale=0.55]{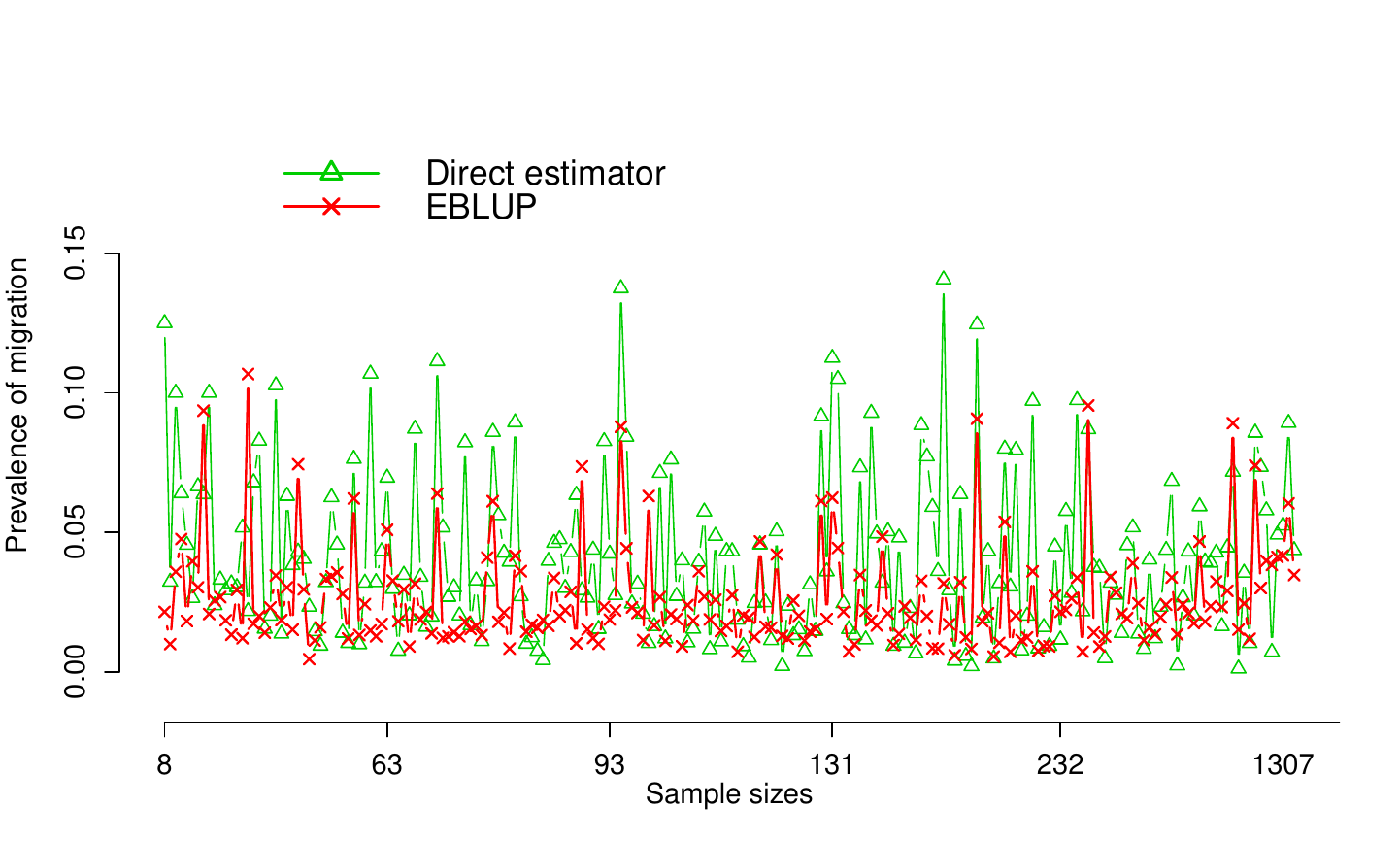}
\end{center}

\vspace*{-6mm}
\caption{Comparison of the estimation of migration through the direct estimator and EBLUP.}
\label{fig:estimators}
\end{figure}

Figure \ref{fig:estimators} shows that the EBLUPs provide more conservative and smooth estimations of prevalences than those obtained using the direct estimators. The EBLUPs of international migration prevalences are comparable with those obtained by using the direct estimator, but they have smaller mean squared errors as shown in Table \ref{tab:tablam2}.

Table \ref{tab:tablam2} presents the estimates of international migration prevalences at the municipal level for large capital cities and border areas.
These are the municipalities where the flow of migrants is of greater interest for studying the population dynamics in Colombia.
Columns 2 and 4 contain the direct and EBLUP estimation, respectively. Columns 3 and 5 contain the estimated coefficients of variation for direct estimators and EBLUPs, respectively.
Table \ref{tab:tablam2} shows that EBLUPs have lower estimated coefficients of variation than direct estimators in all cases.
Further, the estimated coefficients of variation are lower than 20\%. This has considerable practical implications for the final publication of results and for decision making.

\begin{table}[ht]
\renewcommand{\arraystretch}{0.95}
\centering
\begin{tabular}{rlrrrr}
   \hline
 &  Municipality	& Dir	& ecv-Dir(\%)	& EBLUP	& ecv-EBLUP(\%)\\
   \hline

 & Bogot\'a D.C. & 0.04 & 14.75 & 0.03 & 13.22 \\
 & Cali & 0.09 & 14.11 & 0.06 & 10.53 \\
& Armenia & 0.09 & 14.23 & 0.07 & 9.57 \\
 & Cartagena de Indias & 0.07 & 16.80 & 0.03 & 14.47 \\
 & Soledad & 0.04 & 24.79 & 0.02 & 16.74 \\
 & Ch\'ia & 0.05 & 75.30 & 0.04 & 13.87 \\
 & Villa del Rosario & 0.05 & 48.87 & 0.03 & 9.82 \\
  & Manaure & 0.04 & 50.62 & 0.03 & 17.29 \\
  & Valle del Guamuez & 0.05 & 67.01 & 0.03 & 16.69 \\
  & San andr\'es de Tumaco & 0.04 & 45.94 & 0.02 & 20.08 \\
   \hline
\end{tabular}
\caption{Estimates of international migration prevalences at the municipal level.}
\label{tab:tablam2}
\end{table}

\begin{figure}[ht]
\begin{center}
\vspace*{-6mm}

\includegraphics[scale=0.7]{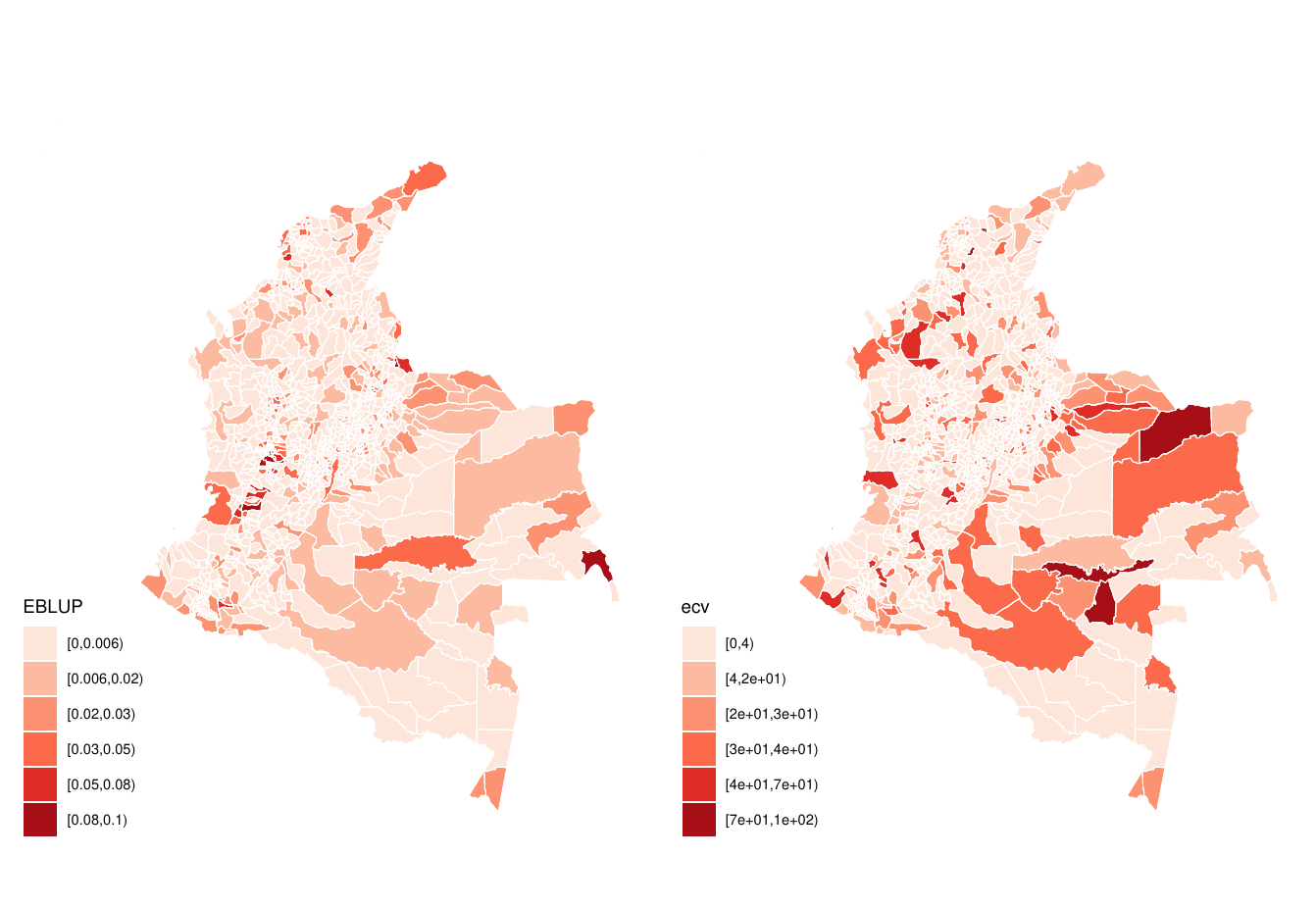}

\end{center}

\vspace*{-6mm}
\caption{Migration prevalence using the EBLUP estimator (left) and estimated coefficients of variation in percentage (right).}
\label{fig:mapa1}
\end{figure}

Figure \ref{fig:mapa1} plots the results for each geographic area and illustrates how in large capital cities and border zones the phenomenon of migration abroad is greater, which is consistent with population dynamics in Colombia.

\section{Concluding remarks}\label{ssec:cuatro}
The main Colombian cities, which are Bogotá, Cali, Medellín, Barranquilla and Cartagena, present some of the highest estimations in the proportions of households that have experienced international migration (with family members abroad). This is interesting inasmuch as these cities are the main areas where services, education and work are concentrated. It is likely that for many international migrants passing through capital cities represents an intermediate phase to look for opportunities to emigrate, whereas for those who are from said municipalities it may be an attractive option to complement their educational and career projects or to search for a better quality of life. Moreover, high indicators can also be seen in municipalities of the coffee-growing region which doubtless have a series of migration determinants and are different to those of the capital cities. This is a topic that merits further investigation in greater detail in order to better identify and characterize these differences.

Likewise, it must be highlighted that events that occur in specific geographical locations, in this case capital cities, often have consequences on their direct neighbors – the adjoining municipalities. In this sense, what was mentioned earlier can be reconfirmed: migratory phenomena do not occur homogeneously within the country. Instead, there is a selection filter meaning that there is a greater prevalence of international emigration in some populations; for example, educational level and employment on an individual scale, but also at the scale of small areas. In fact, it is for these very same reasons that migration is affected by the economic cycle of countries and regions. Situations of economic crisis, such as the current situation in Venezuela, determine changes in the volumes of emigrants and immigrants, taking into account that individuals and families tend to look for the best opportunities in the areas identified.

The result obtained is also relevant insofar as, in Colombia, high-quality disaggregated indicators on a municipal level are only available after each population census, that is, every 10 or 15 years. In the intervening years, it is normal for changes to occur in migration patterns and in the intensity of the phenomenon. For this reason, sample surveys provide data that makes it possible to obtain estimates for interim periods – intercensal estimations – and for periods following censuses, whereby survey results are available – postcensal periods. Additionally, the possibility of including variables that could be catalogued as migration determinants in the models had not been addressed in other demographic sources and applications consulted in the DANE. Hence, it can be affirmed that small areas estimation is very useful for demographic analysis, since estimations which are more-or-less robust and very detailed can be obtained without the need of increasing sample sizes.

The estimation of migration prevalence in Colombian municipalities in intercensal periods is of great importance due to the variability of the migratory flow over recent years. Direct estimations using surveys such as DHS make it possible to obtain estimates at the municipal level, with the disadvantage that these estimates display considerable sample errors. Therefore, a methodology was conducted to estimate migration prevalence at the municipal level using, for the first time, SAE methods, particularly estimation based on Fay-Herriot area models.

In order to estimate migration prevalence in intercensal periods, auxiliary variables at the municipal level were used. They were obtained from previous census studies and applied to a Fay-Herriot model and to the appropriate modeling of variance at the municipal level.  By applying SAE methods, prevalence estimates were obtained with estimated coefficients of variation lower than 20\%, enabling the results to be applied in statistical reports for decision making. The migration prevalence estimations are consistent with the population dynamics of Colombia, and they are similar to estimates achieved using the direct estimator, however with a considerably lower estimation error. Finally, the methodology proposed can be considered for the estimation of other indicators in a future work.

\subsubsection*{Acknowledgements}

This project was supported in part by the Bloomberg Philanthropies Data for Health Initiative,
National Administrative Department of Statistics and Vital Strategies and, by the Spanish grant MTM2015-64842-P.  We thank Jorge Martinez (CEPAL) for his valuable comments on this analysis.

\subsubsection*{References}
\begin{description}
\item
Anand. S., Sen A. (1997). Concepts or human development and poverty! a multidimensional perspective.
{\it United Nations Development Programme, Poverty and human development: Human development papers}, 1–20.
\item
Benavent, R., Morales, D. (2016). Multivariate Fay–Herriot models for small area estimation.
{\it Computational Statistics and Data Analysis}, 94, 372–390.
\item
Burgard, J.P. , Esteban, M.D., Morales, D., Pérez A. (2019). A Fay-Herriot model when auxiliary variables are measured with error.
To appear in {\it TEST}.
\item
Datta, G.S., Lahiri, P. (2000). A unified measure of uncertainty of estimated best linear unbiased predictors
in small area estimation problems. {\it Statistica Sinica}, 10, 613-627.
\item
Edmonston, B., Michalowski, M. (2004). International Migration. In: Shryock, H. Siegel, J. Swanson, D. (eds.): The Methods and Materials of Demography, Second Edition. New York: Academic Press.
\item
Esteban, M.D., Morales, D., P\'erez, A., Santamar\'{\i}a, L. (2012b).
Small area estimation of poverty proportions under area-level time models.
{\it Computational Statistics and Data Analysis}, 56, 2840-2855.
\item
Fay, R.E., Herriot, R.A. (1979). Estimates of income for small places: an application of James-Stein procedures to census data.
\textit{Journal of the American Statistical Association}, 74, 269-277.
\item
Gonz\'alez-Manteiga, W., Lombard\'{\i}a, M. J., Molina, I., Morales, D., Santamar\'{\i}a, L.  (2010).
Small area estimation under Fay-Herriot models with nonparametric estimation of heteroscedasticity.
{\it Statistical Modelling}, 10, 2, 215-239.
\item
H\'ajek, J. (1971) Comment on ``An Essay on the Logical Foundations of Survey Sampling, Part One". In: The Foundations of Survey Sampling, Godambe, V.P. and Sprott, D.A. eds., 236, Holt, Rinehart, and Winston.
\item
Migración-Colombia (2015). Comportamiento migratorio colombiano.
{\it Boletín anual de estadísticas. Enero a diciembre 2015. Migración. Ministerio de Relaciones Exteriores}, 16–23.
URL http://migracioncolombia.gov.co/index.php/es/.
\item[]
Morales, D., Pagliarella, M.C., Salvatore, R. (2015). Small area estimation of poverty indicators under partitioned area-level time models. {\it SORT-Statistics and Operations Research Transactions}, 39, 1, 19-34.
\item
Plane, D.A. (1982). An Information Theoretic Approach to the Estimation of Migration Flows. {\it Journal of Regional Science}, 22,  441-456.
\item
Prasad, N.G.N., Rao, J.N.K. (1990). The estimation of the mean squared error of small-area estimators.
{\it Journal of the American Statistical Association}, 85, 163-171.
\item
Profamilia (2015). The estimation of the mean squared error of small-area estimators.
{\it Encuesta Nacional de Demografía y Salud}.
\item
Rao, J.N.K., Molina, I. (2015). {\it Small area estimation}, second edition. John Wiley, Hoboken (NJ).
\item
Raymer, J. (2007). The Estimation of International Migration Flows: A General Technique Focused on the Origin-Destination Association Structure.
{\it Environment and Planning A}, 39, 985-995.
\item
Raymer, J. and Rogers, A. (2007). Using Age and Spatial Flow Structures in the Indirect Estimation of Migration Streams.
{\it Demography}, 44, 199-223.
\item
Schall, R. (1991). Estimation in generalized linear models with random effects.
{\it Biometrika}, 78, 4, 719–727.
\item
Smith, S.K., Tayman, J., Swanson, D.A. (2013). {\it A practitioner’s guide to state and local population projections}. Springer-Verlag.
\item[]
S\"arndal, C.E., Swensson, B. and Wretman J. (1992). {\it Model assisted survey sampling}. Springer-Verlag.
\item
Willekens, F. (2008). Models of Migration: Observations and Judgements. In {\it International Migration in Europe : Data, Models and Estimates}, 117-147, edited by J. Raymer, J. and Willekens, F. University of Southampton. Division of Social Statistics. John Wiley, Hoboken, New Jersey.
\end{description}

\end{document}